# Exploring Teachers' Perceptions of ChatGPT Through Prompt Engineering

**Dimitrios Gousopoulos**

Adjunct instructor, Panteion University

d.gousopoulos@panteion.gr

**Abstract**

Artificial Intelligence and especially Large Language Models (LLM), such as ChatGPT has revolutionized the way educators work. The results we get from LLMs depend on how we ask them to help us. The process and the technique behind an effective input is called prompt engineering. The aim of this study is to investigate whether science educators in secondary education improve their attitude toward ChatGPT as a learning assistant after appropriate training in prompt engineering. The results of the pilot study presented in this paper show an improvement in the previously mentioned teachers' perceptions.

**Introduction**

For a long time, when people spoke about Artificial Intelligence (AI), they referred to the idea that an AI program could be trained to predict sequences of things. For example, when you watch a movie on Netflix, an underlying algorithm predicts which other movie you might want to see. However, even with this type of artificial intelligence, natural language processing seemed extremely complex and difficult. Eventually, in 2017, scientists introduced the transformer mechanism with attention, which laid the foundations for modern Large Language Models (LLMs) (Vaswani, A., et al., 2017).

This mechanism enabled AI models to pay attention to the entire context surrounding a word and then predict the next word in a sequence. For instance, if we start a sentence by typing "the best place for a trip is," the AI model will use all the knowledge it has learned from its training on large amounts of linguistic data to predict the next most fitting word. Thus, the next word might be "anywhere" with a probability of 60%, "nature" with 30%, and "wonderland" with 10%. The model usually selects the word with the highest probability, but this is not always the case. There is a degree of randomness in these systems, so the second or third most probable word might be chosen instead (Suleyman, M., 2023).

Large Language Models (LLMs) are not limited to generating text; they can also compose music, create games, play chess, solve mathematical problems, generate stunning images from text descriptions, and help software engineers write professional-level code. Therefore, these systems, which are designed to create new content—such as text, images, audio, code, and more—based on training data and in response to prompts, are called Generative Artificial Intelligence programs (GenAI) (Wang et al., 2024).

Artificial intelligence, and especially Large Language Models (LLMs) such as ChatGPT, are bringing rapid changes to the field of science education. Through appropriate prompting, ChatGPT can enhance personalized learning, students' problem-solving skills, and their understanding of physics concepts (Mahligawati et al., 2023; Yeadon & Hardy, 2024; Mustofa et al., 2024). More specifically, ChatGPT can enrich students' learning experience in various ways, thereby improving their learning outcomes and academic performance. As Zhu et al. (2023) emphasized, ChatGPT offers multiple opportunities to support the four types of interactions essential for effective learning: (1) interaction with learning content (e.g., materials), (2) interaction with others (e.g., teachers, peers), (3) interaction with oneself (e.g., self-reflection and self-regulation), and (4) interaction with problem-solving tasks (Gousopoulos, 2025) (e.g., upgrading knowledge and applying it in contexts).

ChatGPT can support both the cognitive and emotional aspects of students' learning experience by offering a personalized learning environment (Kim et al., 2021). Due to its ability to facilitate interactive learning and accommodate individual learning needs, student engagement with ChatGPT can maximize their abilities and motivation to learn, significantly improving academic performance (Murad et al., 2023). The personalized instruction and feedback provided by ChatGPT in any subject or assignment help students receive timely guidance to understand complex problems or concepts, experiment with practical topics, and practice new concepts through a set of tailored exercises (Luo et al., 2023). Since ChatGPT can understand and respond to questions in many languages, it offers a great advantage by allowing students to receive instant guidance on any query (Mhlanga, 2023). As a "teacher," ChatGPT can facilitate a wide range of skills, such as writing, reading, problem-solving in mathematics or physics, and programming practice (Kasneci et al., 2023). It can also assist students by recommending useful learning materials such as practice sheets, manuals, videos, or podcasts aligned with their personal needs or preferences (Murad et al., 2023). By adapting to these cognitive needs, ChatGPT also supports the emotional dimension of learning, as it has been reported to offer a more favorable learning experience, a more comfortable environment, increased self-confidence, and reduced learning anxiety (Adiguzel et al., 2023).

Large Language Models can also assist in educational design, the introduction of innovative teaching methods, and provide feedback to students based on their responses (Mustofa et al., 2024). Specifically, many studies on AI-enhanced learning environments emphasize that the best results can only be achieved through effective human-AI collaboration and that human educators are expected to enrich this experience with new goals, perspectives, actions, and decisions (Jeon & Lee, 2023). Technological innovations such as ChatGPT have already greatly facilitated the transmission and acquisition of knowledge, transforming the teacher's role from a mere transmitter of information to that of a guide and facilitator, enhancing students' skills in accessing knowledge sources, critically evaluating information, and forming a holistic understanding of the world through synthesis (Tsai, 2023; Xiao, 2023). ChatGPT can also serve as a valuable tool for teachers in managing tasks such as creating teaching materials, designing lessons, assessing students, and grading (Mondal et al., 2023). Previous research has shown that ChatGPT is capable of producing comprehensive and creative lesson plans (Farrokhnia et al., 2023; Karakose et al., 2023; Whalen & Mouza, 2023), creating a variety of classroom presentation slides with texts tailored to students' age or proficiency level (Herft, 2023; Mondal et al., 2023), facilitating teaching by offering new ideas and strategies (Luo et al., 2023), and helping design personalized materials and content according to the needs of the course or students (Mondal et al., 2023).

Despite all these advantages, there are certain limitations that educators must take seriously. Such limitations include LLMs' difficulty in handling highly complex content, the lack of visual assistance, and the need for human intervention to ensure the accuracy and correctness of the models' responses (Mustofa et al., 2024). The performance of LLMs varies across academic levels; specifically, they produce more accurate answers to basic physics questions than to university-level ones (Yeadon & Hardy, 2023).

Regarding ChatGPT's use in education, studies have shown that while it can contribute to improving students' critical thinking skills (Bitzenbauer, 2023), it has also been associated with lower academic performance and reduced autonomous learning (Forero & Herrera-Suarez, 2023). Moreover, ChatGPT can be used as a learning assistant, helping teachers design experiments that address students' alternative conceptions of physical phenomena (Gousopoulos, 2024; Kotsis, 2024). However, its use should be approached with caution, and the answers it provides should always be verified by the responsible educator. Finally,

teachers—especially those in the sciences—tend to hold a positive attitude toward using ChatGPT in the educational process (Beege et al., 2024; Gousopoulos, 2024).

One way to make the most of the capabilities of LLMs like ChatGPT is through prompt engineering, a process of designing and refining prompts given to an LLM to elicit higher-quality and more targeted responses (Schmidt et al., 2023). In an educational context, prompt engineering can enhance the learning experience by tailoring the LLM's responses to the specific needs of each class and student. In this way, it further cultivates critical thinking and personalized learning (Leung, 2024).

**Methodology**

This study seeks to investigate the potential change in the opinions of science teachers serving in secondary education regarding the use of ChatGPT as an educational assistant after their training in prompt engineering techniques. The research hypothesis to be tested is as follows: "Training secondary education science teachers in prompt engineering techniques improves their perception of ChatGPT as an educational assistant." The term more positive perception refers both to the improvement of teachers' understanding of the benefits of Artificial Intelligence (AI) in science education and to an increased intention to integrate AI tools into the teaching and learning process.

The participants in this pilot study were 14 teachers from the field of Natural Sciences, working in public and private junior and senior high schools. The selection of participants was based on convenience sampling, with half of the teachers coming from private and the other half from public schools, all of whom were available for training in prompt engineering techniques. The research data were collected using a digital questionnaire, which was based on the study by Nazaretsky et al. (2022) and translated and adapted into Greek. The questionnaire consisted of 14 five-point Likert scale questions covering two thematic areas:

(a) awareness of the benefits of Artificial Intelligence in an educational environment, and

(b) collaboration with Artificial Intelligence to improve the teaching process.

The analysis of responses was carried out using the statistical software R.

The teachers completed the questionnaire prior to their training in prompt engineering techniques. The training, both theoretical and practical, lasted for three weeks (two two-hour sessions per week). Specifically, the teachers were trained in the GPEI methodology (Goal Prompt Evaluation Iteration) (Velásquez-Henao et al., 2023), as well as in a set of prompting techniques such as The Persona Pattern, The Interview Pattern, and Chain-of-Thought. Upon completion of the aforementioned training, the teachers filled out the same initial questionnaire again.

**Results**

For each questionnaire item, the mean value of the responses was calculated, and subsequently, the non-parametric Wilcoxon signed-rank test was applied to examine whether there was a statistically significant difference in the median of the participants' mean scores before and after their training in prompt engineering techniques. The results indicated that there were statistically significant differences ($Z = -3.296$, $p < .001$). It was also observed that, following the implementation of the training, the median of the participants' mean scores on the questionnaire assessing teachers' attitudes toward the use of ChatGPT as an educational assistant was higher than the corresponding median prior to the training.

*Figure 1. Comparison of Pre- and Post-Test*

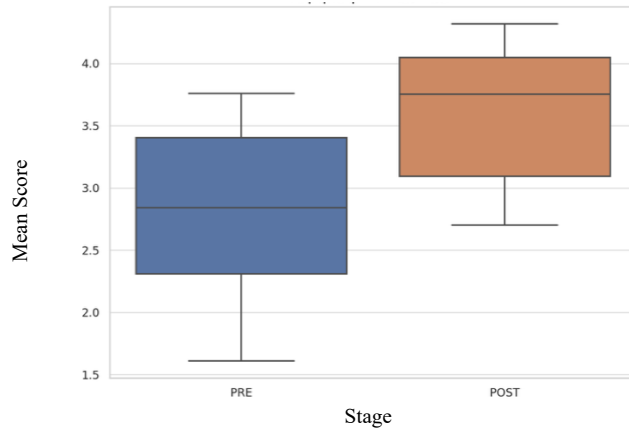

To examine the magnitude and accuracy of the difference between the pre- and post-tests, the data were further analyzed using the bootstrapping method. The reason for applying this method was the small number of participants in the study. Specifically, 10,000 resamples with replacement were performed. For each resample, the difference between the mean scores before and after the intervention (the index of interest) was calculated. The values of this index across all resamples formed an empirical distribution, from which a 95% confidence interval was estimated using the percentile method.

The results of this analysis showed that the confidence interval for the mean difference between the pre- and post-intervention scores ranged from 0.682 to 0.909. The fact that this interval does not include zero indicates a statistically significant change following the intervention.

Therefore, we conclude that there is evidence suggesting that training secondary education teachers in prompt engineering techniques improves their perception of ChatGPT as an educational assistant, thus supporting our initial research hypothesis.

*Figure 2. Distribution of Difference (Bootstrapping)*

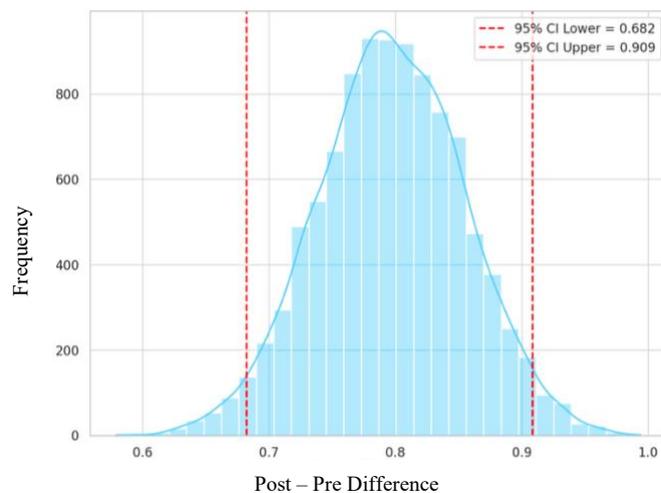

## Discussion

Large Language Models (LLMs) have firmly entered our daily lives and have significantly transformed the way we process and interact with information. Despite the general skepticism surrounding artificial intelligence, it is important to recognize that AI is not meant to replace the role of the teacher but rather to assist them with some of the repetitive tasks they perform daily, providing the necessary time to think, plan, and implement innovative educational approaches for the benefit of their students.

The goal of this pilot study was to investigate whether training science teachers in prompt engineering—that is, the technique of how to interact effectively with LLMs—improves their perception of ChatGPT as an educational assistant. The results indeed showed an improvement in teachers' attitudes: following the instructional intervention, teachers demonstrated greater confidence in what they could achieve with an LLM and increased appreciation for such systems as educational assistants.

Furthermore, the findings suggest that teacher training in prompt engineering techniques can have practical applications in the educational process. For example, teachers can use prompting techniques such as the Persona Pattern to design personalized learning scenarios, develop inquiry-based activities in physics with the help of ChatGPT, and identify the learning difficulties their students face. They can also employ the tool to generate assessment questions tailored to the level and needs of each class. This practice supports differentiated instruction and promotes inclusive learning (Gousopoulos,2023; Gousopoulos 2024).

It is worth noting, however, that despite the positive findings of this pilot study, the limited number of participants calls for a more cautious interpretation of the results. Additionally, the questionnaire used was not tested for validity and reliability in the Greek context due to time constraints. The duration of the training intervention was also relatively short, which should be considered when interpreting the outcomes.

Finally, future research would benefit from incorporating a qualitative approach to further explore the factors influencing the change in teachers' attitudes toward the use of ChatGPT in education.